\documentclass[aps,prb,twocolumn,superscriptaddress,a4paper]{revtex4-1}
\usepackage[dvipdfmx]{graphicx}

\usepackage{amsmath,amsthm,amssymb}
\usepackage{bm}
\usepackage{color}
\usepackage{ifthen}
\usepackage[svgnames,psnames]{xcolor}
\usepackage{multirow,bigdelim}

\newcommand{\1}{\mbox{1}\hspace{-0.25em}\mbox{l}}

\newcommand{\ii}{\mathrm{i}}

\begin{document}

\title{
Exceptional points and non-Hermitian skin effects under nonlinearity of eigenvalues
}
\author{
Tsuneya Yoshida
}
\affiliation{%
Department of Physics, Kyoto University, Kyoto 606-8502, Japan
}%
\affiliation{
Institute for Theoretical Physics, ETH Zurich, 8093 Zurich, Switzerland
}

\author{
Takuma Isobe
}
\author{
Yasuhiro Hatsugai
}
\affiliation{
Department of Physics, University of Tsukuba, Ibaraki 305-8571, Japan
}

\date{\today}
\begin{abstract}
Band structures of metamaterials described by a nonlinear eigenvalue problem are beyond the existing topological band theory.
In this paper, we analyze non-Hermitian topology under the nonlinearity of eigenvalues.
Specifically, we elucidate that such nonlinear systems may exhibit exceptional points and non-Hermitian skin effects which are unique non-Hermitian topological phenomena. The robustness of these non-Hermitian phenomena is clarified by introducing the topological invariants under nonlinearity which reproduce the existing ones in linear systems.
Furthermore, our analysis elucidates that exceptional points may emerge even for systems without an internal degree of freedom where the equation is single component.
These nonlinearity-induced exceptional points are observed in mechanical metamaterials, e.g., the Kapitza pendulum.
\end{abstract}
\maketitle

\section{Introduction}
Topology of quantum states is one of the central issues of condensed matter physics~\cite{TI_review_Hasan10,TI_review_Qi10,Kane_2DZ2_PRL05,Kane_Z2TI_PRL05_2,Qi_TFT_PRB08,Kitaev_chain_01,Thouless_PRL1982,Halperin_PRB82,Hatsugai_PRL93} because of robust band structures which are origins of characteristic transport~\cite{Klitzing_IQHE_PRL1980,Thouless_PRL1982,Qi_TFT_PRB08} properties and exotic particles~\cite{Majorana_Mourik,Majorana_Rokhinson2012,Majorana_Anindya2012}.
In these years, platforms of topological phenomena are extended to open quantum systems~\cite{Partanen_EPQbit_PRB2019,Naghiloo_EPQbit_NatPhys2019,Takasu_nHPTcoldAtom_PTEP2020,TELeePRL16_Half_quantized,ZPGong_PRL17,Bergholtz_Review19,Ashida_nHReview_AdvPhys,Gong_class_PRX18,KKawabata_TopoUni_NatComm19,Kawabata_gapped_PRX19,Zhou_gapped_class_PRB19}, elucidating that the interplay between topology and non-Hermiticity exhibits unique topological phenomena
that do not have Hermitian counterparts~\cite{Hatano_PRL96,Hatano_nHSkin_PRB1997,Esaki_nH_PRB11,Hu_nH_PRB11,Alvarez_nHSkin_PRB18,SYao_nHSkin-2D_PRL18,KFlore_nHSkin_PRL18,EElizabet_PRBnHSkinHOTI_PRB19,Yokomizo_BBC_PRL19,Okuma_EdgeInst_PRL2019,kawabata_NBlochBBC_PRB2020}.
One of the representative examples is the exceptional points on which band touching occurs for both real and imaginary parts~\cite{Rotter_EP_JPA09,Berry_EP_CzeJPhys2004,Heiss_EP_JPA12,HShen2017_non-Hermi,VKozii_nH_arXiv17,Zyuzin_nHEP_PRB18,Yoshida_EP_DMFT_PRB18,Shen_EPQosci_PRL2018,Budich_SPERs_PRB19,Okugawa_SPERs_PRB19,Yoshida_SPERs_PRB19,Zhou_SPERs_Optica19,Kawabata_gapless_PRL19,Yoshida_nHReview_PTEP20,Konig_BraidEP3PRR2023}, leading to the breakdown of diagonalizability of the Hamiltonian. A non-Hermitian skin effect is another unique non-Hermitian phenomenon that represents the extreme sensitivity of eigenvalues and eigenstates to boundary conditions~\cite{SYao_nHSkin-1D_PRL18,Xiao_nHSkin_Exp_NatPhys20,Lee_Skin19,Okuma_BECskin19,Zhang_BECskin19,Borgnia_ptGapPRL2020,Okugawa_HOSE_PRB2020,Kawabata_HOSE_PRB2020,Hwang_JuncSkinPRB2023,Song_LSkin_PRL2019,Haga_LSkin_PRL2021,Okuma_NHReview_AnnRevCondMatt2023,Lin_NHSEReview2023}.  
Non-Hermitian topology results in an extensive number of eigenstates localized around one of the edges~\cite{Okuma_BECskin19,Zhang_BECskin19} which are called skin modes.

The platform of the above topological phenomena is further extended to a wide range of classical systems from metamaterials~\cite{Haldane_chiralPHC_PRL08,Raghu_chiralPHC_PRA08,Wang_chiralPHC_Nature09,Fu_chiralPHCapp_AppPhys10,KTakata_EP_PRL2018,Ozawa_TopoPhoto_RMP19,Harari_TopoLaser_Science18,Banders_TopoLaser_Science18,ProdanPRL09,Kane_NatPhys13,Kariyado_SR15,Susstrunk_TopoMech_Sci15,Suesstrunk_Mech-class_PNAS16,Yoshida_SPERs_mech19,Scheibner_MechSkin_PRL2020,Albert_Topoelecircit_PRL15,Ningyuan_Topoelecircit_PRX15,Victor_Topoelecircit_PRL15,Lee_Topoelecircit_CommPhys18,Hofmann_EleCirChern_PRL19,Helbig_ExpSkin_19,Hofmann_ExpRecipSkin_19,Yoshida_MSkinPRR20,Yoshida_topodiff_SciRep20,Hu_TopoDiff_AdvMat2022,Delplace_topoEq_Science17,Delplace_Resul_PRL21}
 to biological systems
~\cite{Sone_TopoActMattPRL2019,Yamauchi_ActiveMatter_arXiv20,Sone_nHTopoActMattNatComm2020,Knebel_RPSchain_PRL20,Yoshida_chiralRPS_PRE21,Yoshida_nHgame_SciRep2022}.
The topological perspective on these classical systems relies on the fact that the linear equations describing these systems are written as eigenvalue problems, providing a mathematical analogy to the Schr\"odinger equation. 
For instance, Maxwell equations written as an eigenvalue problem elucidate the topological origin of chiral edge modes of electromagnetic fields in a photonic crystal~\cite{Haldane_chiralPHC_PRL08,Raghu_chiralPHC_PRA08}.
While the topological band theory is applied to a variety of linear systems,
classical systems may include two types of nonlinearity, nonlinearity of eigenvalues~\cite{Isobe_NLomegaTopo_PRL2024} and that of eigenstates~\cite{Sone_NLpsiTopo_NatPhys2024,Schindler_NLpsiTopo_NatPhys2024}.

The existence of such nonlinear classical systems leads to intriguing issues: topological phenomena under nonlinearity.
For each type of nonlinearity, Hermitian topology is analyzed, elucidating the robustness of edge states~\cite{Sone_NLpsiTopo_NatPhys2024,Isobe_NLomegaTopo_PRL2024}.
However, non-Hermitian topology under nonlinearity remains elusive. In particular, 
the fate of exceptional points and non-Hermitian skin effects 
under nonlinearity remains a crucial question to be addressed.

In this paper, we analyze non-Hermitian topology under the nonlinearity of eigenvalues. Specifically, we elucidate that exceptional points and non-Hermitian skin effects, representative non-Hermitian topological phenomena, survive under the nonlinearity of eigenvalues. 
The robustness of these phenomena is clarified by extending the topological invariants to the nonlinear systems. Our analysis also elucidates that exceptional points may emerge for systems without an internal degree of freedom which are described by the equation of single component.
Such nonlinearity-induced exceptional points emerge for mechanical systems (e.g., the Kapitza pendulum) where the exceptional point separates stable and unstable modes.

The rest of this paper is organized as follows.
Section~\ref{sec: rev of NLEGP} provides a brief review of a nonlinear eigenvalue problem.
Section~\ref{sec: EPandSPEP} elucidates the emergence of exceptional points and their symmetry-protected variants under nonlinearity.
Section~\ref{sec: NHSE} elucidates the robustness of the non-Hermitian skin effect under the nonlinearity.
We also demonstrate the emergence of the nonlinearity-induced exceptional points for mechanical metamaterials in Sec.~\ref{sec: EPs in MetMat} which is accompanied by a short summary.
Appendices are devoted to the details of classical metamaterials with nonlinearity of eigenvalues.

\section{Nonlinear eigenvalue problem}
\label{sec: rev of NLEGP}
We consider a nonlinear eigenvalue problem~\cite{Guttel_NLEVP_ActNum2017} defined in the momentum space 
\begin{eqnarray}
\label{eq: NLEVP}
F(\omega,\bm{k}) |\psi(\omega,\bm{k}) \rangle
 &=&0,
\end{eqnarray}
with $N\times N$ matrix $F(\omega,\bm{k})$, eigenvalues $\omega \in \mathbb{C}$, and right eigenvectors $|\psi \rangle$ having $N$ components.
Vector $\bm{k}$ specifies the momentum (i.e., wavenumber).
Eigenvalues $\omega_n(\bm{k})\in \mathbb{C}$ ($n=1,2,\ldots$) are specified by $\det F(\omega,\bm{k})=0$.

Here we define that the point-gap opens for a given $\omega_{\mathrm{ref}}\in\mathbb{C}$
when no eigenvalue equals to $\omega_{\mathrm{ref}}$
[i.e., $\mathrm{det}F(\omega_{\mathrm{ref}},\bm{k}) \neq 0$] in the momentum space. This definition reproduces the definition of the point-gap in linear systems where the matrix is written as $F(\omega,\bm{k})=H(\bm{k})-\omega\1$ with $\1$ being the $N\times N$ identity matrix~\cite{gaplinear_ftnt}.

The nonlinear eigenvalue problem [Eq.~(\ref{eq: NLEVP})] describes band structures of metamaterials. For instance, the frequency dependence of permittivity and permeability results in the nonlinear eigenvalue problem of electromagnetic fields~\cite{Kuzmiak_NLPhotoPRB1994}~\cite{Chern_NLEVPPhoto_PRE2006}. In addition, the nonlinear eigenvalue problem describes mechanical oscillators with internal structures where the effective mass is frequency-dependent~\cite{Huang_NLEVPMech_IntJ2009,Lee_NLEVPMech_PRB2016}.

\section{
Exceptional points under nonlinearity
}
\label{sec: EPandSPEP}

On an exceptional point, band touching occurs for both real and imaginary parts of eigenvalues which is protected by point-gap topology for linear systems.
We elucidate that exceptional points and their symmetry-protected variants are robust even under nonlinear systems.
In addition, our analysis clarifies that exceptional points may emerge, even when $F(\omega,\bm{k})$ is a $1\times1$ matrix [see Eqs.~(\ref{eq: 1x1 F})~and~(\ref{eq:w+, w- gen})]. The emergence of such nonlinearity-induced exceptional points is in sharp contrast to linear systems.

\subsection{Exceptional points in two dimensions}
We consider a two-dimensional nonlinear system described by $1 \times 1$ matrix
\begin{eqnarray}
\label{eq: 1x1 F}
F(\omega,\bm{k}) &=&\omega^2-2a\omega+b(\bm{k})
\end{eqnarray}
with $a\in \mathbb{R}$ and $b(\bm{k})=k_x+\ii k_y$. Wavenumber in the $x$ and $y$ direction are denoted by $k_x$ and $k_y$.
Factorizing Eq.~(\ref{eq: 1x1 F}), we obtain
\begin{eqnarray}
\label{eq: F=(w-w+)(w-w-)}
F(\omega,\bm{k})
&=&
[\omega-\omega_+(\bm{k})][\omega-\omega_-(\bm{k})]
\end{eqnarray}
with
\begin{eqnarray}
\label{eq:w+, w- gen}
\omega_\pm(\bm{k})
&=&
a\pm \sqrt{a^2-b(\bm{k})}.
\end{eqnarray}
Equation~(\ref{eq:w+, w- gen}) elucidates the emergence of a nonlinearity-induced exceptional point~\cite{EPLinear_ftnt}; two bands touch at $\bm{k}_{\mathrm{EP}}=(a^2,0)$ and $\omega_{\mathrm{ref}}=a$.

For topological characterization of the exceptional point, we introduce the winding number for nonlinear systems 
\begin{eqnarray}
\label{eq: NL W}
W(\omega_{\mathrm{ref}})
&=&
\oint \frac{d\bm{k}}{2\pi i} \cdot \bm{\nabla}_{\bm{k}} \mathrm{Arg}[\mathrm{det} F(\omega_{\mathrm{ref}},\bm{k})]
\end{eqnarray}
with $\bm{\nabla}_{\bm{k}}=(\partial_{k_x},\partial_{k_y})$. Derivative with respect to $k_\mu$ ($\mu=x,y$) is denoted by $\partial_{k_\mu}$. The integral is taken over a closed loop enclosing the exceptional point on which band touching occurs at $\omega_{\mathrm{ref}}\in \mathbb{C}$. 
The topological invariant $W(\omega_{\mathrm{ref}})$ counts how many times the eigenvalues $\omega_\pm(\bm{k})$ wind around the point $\omega_{\mathrm{ref}}$ in the complex plane of $\omega$.
In the case of $F(\omega_{\mathrm{ref}},\bm{k})=H(\bm{k})-\omega_{\mathrm{ref}}\1$, $W(\omega_{\mathrm{ref}})$ is reduced to the winding number in the linear systems characterizing exceptional points~\cite{HShen2017_non-Hermi,Kawabata_gapless_PRL19}.

The nonlinearity-induced exceptional point 
is characterized by $W=1$ with $\omega_{\mathrm{ref}}=a$ which can be seen by substituting $F(a,\delta \bm{k}+\bm{k}_{\mathrm{EP}}) = \delta k_x+\ii \delta k_y$
to Eq.~(\ref{eq: NL W}) where $\delta \bm{k}$ and $\bm{k}_{\mathrm{EP}}$ are $\delta \bm{k}=\delta k_x+\ii \delta k_y$ and $\bm{k}_{\mathrm{EP}}=(a^2,0)$, respectively.
This result is consistent with the fact that the eigenvalues wind the point $\omega_{\mathrm{ref}}=a$ in the counter-clockwise direction once [see Eq.~(\ref{eq: F=(w-w+)(w-w-)})].

Generalizing the above argument of $1\times1$ matrix $F(\omega,\bm{k})$ elucidates how the winding number [Eq.~(\ref{eq: NL W})] characterizes exceptional points.
A key ingredient is that the following relation holds
\begin{eqnarray}
\label{eq: detF=f(w)(w-w+)(w-w-)}
\mathrm{det}F(\omega,\bm{k})
&=&
f(\omega,\bm{k}) [\omega-\omega_1(\bm{k})] [\omega-\omega_2(\bm{k})]
\end{eqnarray}
around an exceptional point where only two bands touch $\omega_1=\omega_2=\omega_{\mathrm{ref}}$ at $\bm{k}_{\mathrm{EP}}$.
Here, we have supposed that $f(\omega,\bm{k})$ is a continuous function that does not have zero around the exceptional point; otherwise, more than three bands touch at this point.
With Eq.~(\ref{eq: detF=f(w)(w-w+)(w-w-)}), we can see that the winding number counts how many times the eigenvalues, $\omega_1(\bm{k})$ and $\omega_2(\bm{k})$, wind around the point $\omega_{\mathrm{ref}}$ in the complex plane of eigenvalues.
We note that the function $f(\omega_{\mathrm{ref}},\bm{k})$ does not contribute to the winding number because it remains non-singular
(i.e., finite) around the exceptional point.

\subsection{
Application of Eq.~(\ref{eq: NL W}) for a nonlinear system of a $2\times 2$ matrix
}
We consider a toy model
\begin{eqnarray}
\label{eq: F toy EP}
F(\omega,\bm{k})
&=&
\left(
\begin{array}{cc}
0 & k_x+\ii k_y \\
\omega+1 & 0
\end{array}
\right)
+\omega \tanh(\omega)\sigma_3-\omega\sigma_0 \nonumber\\
\end{eqnarray}
with Pauli matrices $\sigma_i$ ($i=1,2,3$) and the $2\times 2$ identity matrix $\sigma_0$.

\begin{figure}[!t]
\begin{minipage}{0.49\hsize}
\begin{center}
\includegraphics[width=1\hsize,clip]{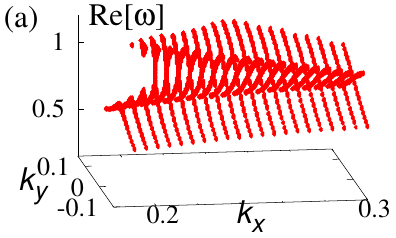}
\end{center}
\end{minipage}
\begin{minipage}{0.49\hsize}
\begin{center}
\includegraphics[width=1\hsize,clip]{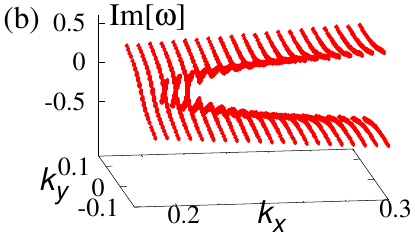}
\end{center}
\end{minipage}
\begin{minipage}{0.49\hsize}
\begin{center}
\includegraphics[width=1\hsize,clip]{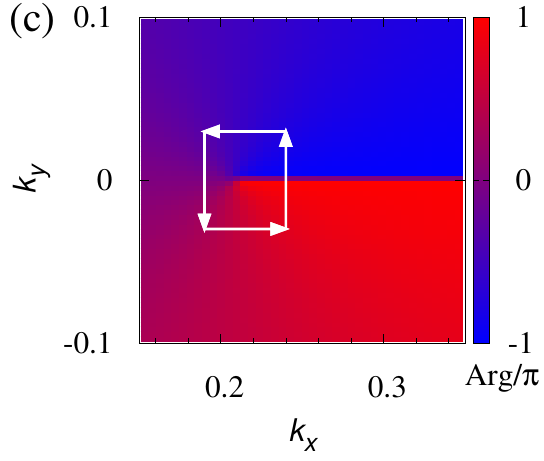}
\end{center}
\end{minipage}
\begin{minipage}{0.49\hsize}
\begin{center}
\includegraphics[width=1\hsize,clip]{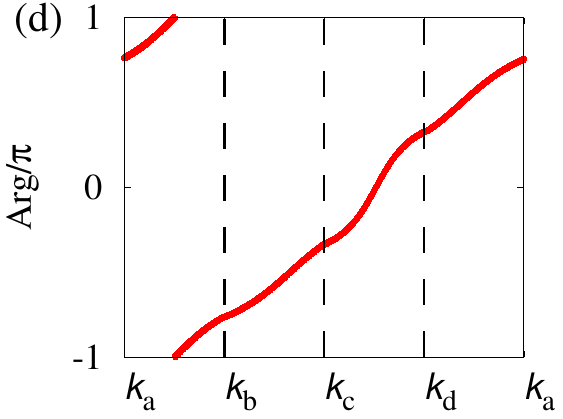}
\end{center}
\end{minipage}
\caption{
(a) [(b)]: Real [imaginary] part of eigenvalues. Two bands touch at $\bm{k}_{\mathrm{EP}}\sim(0.21,0)$.
The eigenvalues $\omega$ and eigenmodes $|\psi(\omega,\bm{k})\rangle$ are numerically obtained which satisfy 
$F(\omega,\bm{k})|\psi(\omega,\bm{k})\rangle=\lambda|\psi(\omega,\bm{k})\rangle$ with $|\lambda|<0.005$.
(c) and (d): Argument of $\mathrm{det}F(\omega_{\mathrm{ref},\bm{k}})$ with $\omega_{\mathrm{ref}}=0.9$.
In panel (d), the argument is plotted along the path illustrated in panel (c).
Here, $\bm{k}$'s are defined as
$\bm{k}_{\mathrm{a}}=(0.24,-0.03)$, 
$\bm{k}_{\mathrm{b}}=(0.24,0.03)$, 
$\bm{k}_{\mathrm{c}}=(0.19,0.03)$, 
$\bm{k}_{\mathrm{d}}=(0.19,-0.03)$.
}
\label{fig: EP toy}
\end{figure}

As shown in Figs.~\ref{fig: EP toy}(a)~and~\ref{fig: EP toy}(b),
band touching occurs at $\bm{k}_{\mathrm{EP}}\sim(0.21,0)$ for both real and imaginary parts, which indicates the emergence of an exceptional point.
This exceptional point is protected by the winding number $W=1$, which can be seen in Figs.~\ref{fig: EP toy}(c)~and~\ref{fig: EP toy}(d). 

\subsection{Symmetry-protected exceptional points in one dimension}

Symmetry constraints may protect exceptional points under nonlinearity. Furthermore, nonlinearity may induce symmetry-protected exceptional points for systems without an internal degree of freedom (i.e., systems described by a $1\times1$ matrix).

Consider a nonlinear eigenvalue problem with a matrix $F$ satisfying 
\begin{eqnarray}
\label{eq: UFom*U=Fom}
F(\omega^*,k)
&=&
U [F(\omega,k)]^\dagger U,
\end{eqnarray}
with a unitary and Hermitian matrix $U$ ($U^2=\1$). Here, symbols ``${}^*$" and ``${}^\dagger$" denote complex conjugate and Hermitian conjugate, respectively.
Because Eq.~(\ref{eq: UFom*U=Fom}) is reduced to pseudo-Hermiticity for $F(\omega,k)=H(k)-\omega \1$, we denote the above symmetry by nonlinear pseudo-Hermiticity.

Under the nonlinear pseudo-Hermiticity~(\ref{eq: UFom*U=Fom}), we have 
\begin{eqnarray}
\mathrm{det}F(\omega^*,k)
&=&
[\mathrm{det}F(\omega,k)]^*,
\end{eqnarray}
which indicates eigenvalues form a pair $\omega_{n'}(k)=\omega^*_n(k)$ ($n\neq n'$) or take real numbers $\omega_{n}(k)\in \mathbb{R}$.
This symmetry constraint protects exceptional points for $\omega_{\mathrm{ref}}\in \mathbb{R}$.

In particular, the above symmetry constraint results in a nonlinearity-induced symmetry-protected exceptional point in one dimension.
For instance, let us consider a $1\times 1$ matrix 
\begin{eqnarray}
 \label{eq: SPEP 1x1 F}
 F(\omega,k) &=&\omega^2-2a\omega+k 
\end{eqnarray}
with $a\in \mathbb{R}$. Matrix $F$ satisfies $F(\omega^*,k)=[F(\omega,k)]^*$ which corresponds to Eq.~(\ref{eq: UFom*U=Fom}).
In this case, a symmetry-protected exceptional point emerges at $k_{\mathrm{EP}}=a^2$ and $\omega_{\mathrm{ref}}=a$; eigenvalues are written as
$\omega=a \pm \sqrt{a^2-k}$ which shows band touching for $k_{\mathrm{EP}}$.

For the characterization of such symmetry-protected exceptional points, we introduce the zeroth Chern number for nonlinear systems
\begin{eqnarray}
\label{eq: N0Ch}
N_{0\mathrm{Ch}}
&=&
\sum_{n} \frac{[1-\mathrm{sgn}(\lambda_{\mathrm{U}n})]}{2}
\end{eqnarray}
with $\mathrm{sgn}(x)$ taking $1$ [$-1$] for $x>0$ [$x<0$]. Here, $\lambda_{\mathrm{U}n}$ ($n=1,2,\ldots$) denotes eigenvalues of Hermitian matrix
\begin{eqnarray}
\label{eq: FU}
F_{U}(\omega_{\mathrm{ref}},k)
&=&
F(\omega_{\mathrm{ref}},k)U
\end{eqnarray}
with $\omega_{\mathrm{ref}}\in \mathbb{R}$.
Namely, $N_{0\mathrm{Ch}}$ counts the number of eigenmodes with negative eigenvalue of $F_{U}(\omega_{\mathrm{ref}},k)$.
In the case of $F(\omega_{\mathrm{ref}},k)=H(k)-\omega_{\mathrm{ref}}\1$, $N_{0\mathrm{Ch}}(\omega_{\mathrm{ref}})$ is reduced to the zeroth Chern number for linear systems~\cite{Shiozaki_TopoClassPntg_PRB2014,Yoshida_SPERs_PRB19}.

The nonlinearity-induced symmetry-protected exceptional point [see Eq.~(\ref{eq: SPEP 1x1 F})] is characterized by $N_{0\mathrm{Ch}}=1$ with $\omega_{\mathrm{ref}}=a$.
For $\omega_{\mathrm{ref}}=a$, we have 
\begin{eqnarray}
\label{eq: 1x1 FU}
F_U(\omega_{\mathrm{ref}},k)=F(\omega_{\mathrm{ref}},k)=-(a^2-k)
\end{eqnarray}
 which corresponds to the discriminant of polynomial function $F(\omega)$ of degree two [see Eq.~(\ref{eq: F=(w-w+)(w-w-)})].
Thus, the zeroth Chern number takes 
$N_{0\mathrm{Ch}}(a)=1$ [$N_{0\mathrm{Ch}}(a)=0$] for $a^2>k$ [$a^2<k$], which demonstrates that the symmetry-protected exceptional point is characterized by the zeroth Chern number [Eq.~(\ref{eq: FU})].

Generalizing the above argument of the $1\times1$ matrix elucidates how the zeroth Chern number [Eq.~(\ref{eq: FU})] characterizes symmetry-protected exceptional points in generic nonlinear systems.
We suppose that two bands touch $\omega_1=\omega_2=\omega_{\mathrm{ref}}\in \mathbb{R}$ at $k=k_{\mathrm{EP}}$.
A key ingredient is $\mathrm{det}F(\omega,k)$ which is written as 
\begin{eqnarray}
\mathrm{det}F(\omega,k)&=&f(\omega,k)D(\omega,k), \\
D(\omega,k) &=& [\omega-\omega_1(k)][\omega-\omega_2(k)],
\end{eqnarray}
around the exceptional point.
Here, $f(\omega,k)$ is a continuous function that does not have zero around the symmetry-protected exceptional point. In particular, $f(\omega_{\mathrm{ref}},k)$ is real because both $\mathrm{det}F(\omega_{\mathrm{ref}},k)$ and $D(\omega_{\mathrm{ref}},k)$ are real. The latter corresponds to the discriminant as discussed in the case of the $1\times 1$ matrix [see Eq.~(\ref{eq: 1x1 FU})].

Here, suppose that the eigenvalues $\omega_1$ and $\omega_2$ satisfy $\omega_1,\omega_2 \in \mathbb{R}$ for $k<k_{\mathrm{EP}}$ [$\omega_1=\omega^*_2$ for $k>k_{\mathrm{EP}}$].
In this case, sign of $D(\omega_{\mathrm{ref}},k)$
changes at $k=k_{\mathrm{EP}}$, which results in sign change of $\mathrm{det}F_U(\omega_{\mathrm{ref}},k)$ because 
$\mathrm{det}F_U(\omega_{\mathrm{ref}},k)=\mathrm{det}{U}\mathrm{det}F(\omega_{\mathrm{ref}},k)$ holds and $f(\omega,k)$ remains finite.
Recalling the relation $\mathrm{det}F_U(\omega_{\mathrm{ref}},k)=(-1)^{N_{0\mathrm{Ch}}}$, we can see that  $N_{0\mathrm{Ch}}$ 
jumps at $k=k_{\mathrm{EP}}$, indicating that zeroth Chern number
characterizes the symmetry-protected exceptional points.

\subsection{
Application of Eq.~(\ref{eq: N0Ch}) for a nonlinear system of a $2\times 2$ matrix
}
We consider a toy model
\begin{eqnarray}
F(\omega,k) &=& 
(k+1)\sigma_1+\omega \tanh(\omega) \sigma_3-\omega \sigma_0
\end{eqnarray}
with Pauli matrices $\sigma_i$ ($i=1,2,3$) and the $2\times 2$ identity matrix $\sigma_0$.
This model preserves the nonlinear pseudo-Hermiticity with $U=\sigma_0$ [see Eq.~(\ref{eq: UFom*U=Fom})].

\begin{figure}[!h]
\begin{minipage}{0.49\hsize}
\begin{center}
\includegraphics[width=1\hsize,clip]{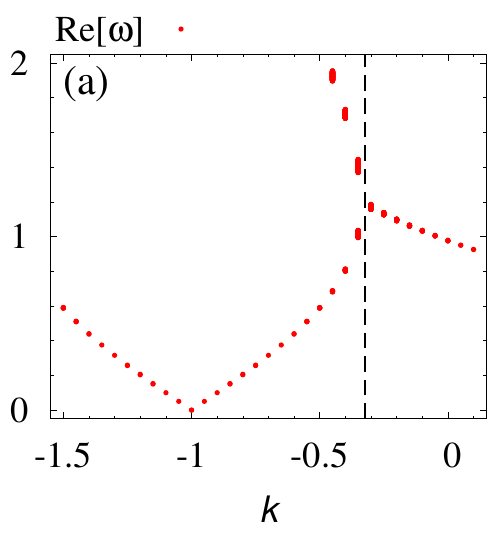}
\end{center}
\end{minipage}
\begin{minipage}{0.49\hsize}
\begin{center}
\includegraphics[width=1\hsize,clip]{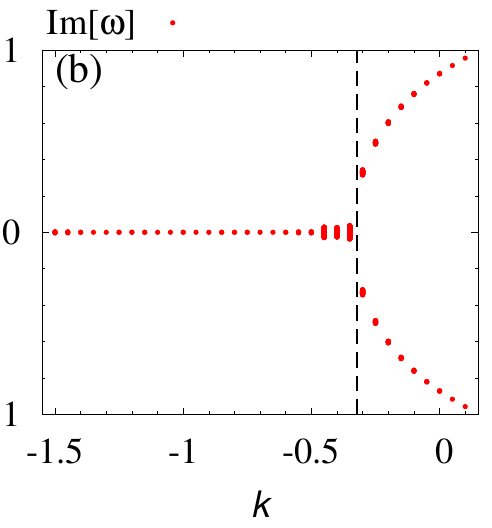}
\end{center}
\end{minipage}
\begin{minipage}{0.7\hsize}
\begin{center}
\includegraphics[width=0.9\hsize,clip]{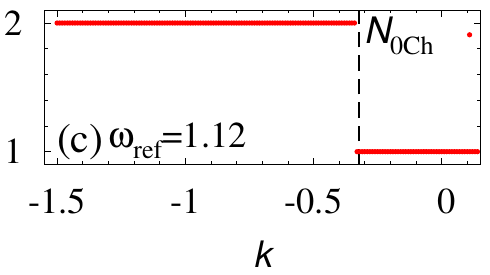}
\end{center}
\end{minipage}
\caption{
(a) [(b)]: Real [imaginary] part of eigenvalues. 
Two bands touch at $k_{\mathrm{EP}}\sim-0.34$.
The numerically obtained
eigenvalues $\omega$ and eigenmodes $|\psi(\omega,\bm{k})\rangle$
satisfy 
$F(\omega,\bm{k})|\psi(\omega,\bm{k})\rangle=\lambda|\psi(\omega,\bm{k})\rangle$ with $|\lambda|<0.002$.
(c): The zeroth Chern number $N_{0\mathrm{Ch}}$ as a function of $k$
for $\omega_{\mathrm{ref}}=1.12$.
The dashed vertical lines denote $k=k_{\mathrm{EP}}$.
}
\label{fig: SPEP toy}
\end{figure}

As shown in Figs.~\ref{fig: SPEP toy}(a)~and~\ref{fig: SPEP toy}(b), this model hosts an symmetry-protected exceptional point at $k_{\mathrm{EP}}=-0.34$. This symmetry-protected exceptional point is protected by the zeroth Chern number $N_{0\mathrm{Ch}}$, which can be seen in Fig.~\ref{fig: SPEP toy}(c).

\section{
non-Hermitian skin effect
}
\label{sec: NHSE}

\begin{figure}[!h]
\begin{minipage}{0.7\hsize}
\begin{center}
\includegraphics[width=1\hsize,clip]{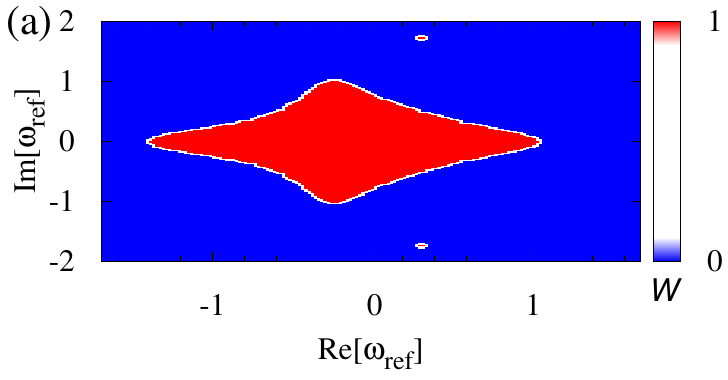}
\end{center}
\end{minipage}
\begin{minipage}{0.49\hsize}
\begin{center}
\includegraphics[width=1\hsize,clip]{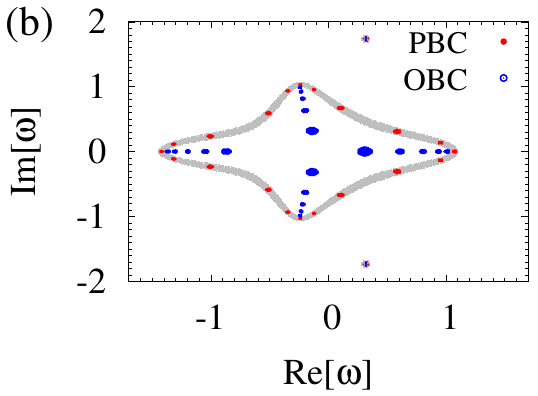}
\end{center}
\end{minipage}
\begin{minipage}{0.49\hsize}
\begin{center}
\includegraphics[width=1\hsize,clip]{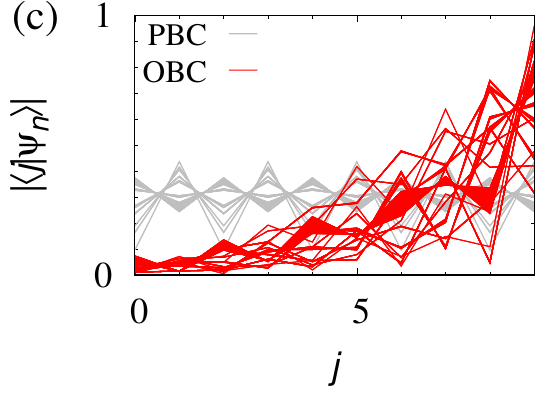}
\end{center}
\end{minipage}
\caption{
Numerical data of a toy model [Eq.~(\ref{eq: NL HatanoNelson})] for $(t_{\mathrm{R}},t_{\mathrm{L}})=(1,0.5)$.
(a): Color-plot of winding number $W$ as a function of the real and imaginary parts of $\omega_{\mathrm{ref}}$.
(b): Eigenvalues under open or periodic boundary conditions. 
The red (blue) dots denote data obtained under periodic (open) boundary conditions for $L=10$.
The gray dots denote data obtained by analyzing the Fourier transformed matrix $F(\omega,k)$.
(c): Eigenmodes under open or periodic boundary conditions. 
The gray (red) lines denote data obtained under periodic (open) boundary conditions.
In pannels~(b)~and~(c), we plot $\omega$ and $|\psi\rangle$ satisfying $ F(\omega)|\psi\rangle=\lambda |\psi\rangle$ with a complex value $|\lambda|<0.05$.
}
\label{fig: NHSE toy}
\end{figure}

A non-Hermitian skin effect, another phenomenon unique to non-Hermitian systems, survives under the nonlinearity of eigenvalues. Namely, nonlinear systems may exhibit extreme sensitivity of eigenvalues and eigenmodes to boundary conditions [see Fig.~\ref{fig: NHSE toy}]. 
In this section, we address characterization of a non-Hermitian skin effect under nonlinearity.

We consider a one-dimensional lattice described by a nonlinear eigenvalue problem [Eq.~(\ref{eq: NLEVP})].
When the winding number $W(\omega_{\mathrm{ref}})$ [Eq.~(\ref{eq: NL W})] is finite under periodic boundary conditions, the system exhibits the skin effect.
This fact can be seen by extending the argument of 
Ref.~\onlinecite{Okuma_BECskin19} 
where the topological origin of the skin effect is elucidated in linear systems.

Firstly, we consider the following Hermitian matrix
\begin{eqnarray}
\tilde{F}(\omega_{\mathrm{ref}},k) &=& 
\left(
\begin{array}{cc}
0 &  F(\omega_{\mathrm{ref}},k) \\
F^\dagger(\omega_{\mathrm{ref}},k) & 0
\end{array}
\right)_\tau
\end{eqnarray}
which describes a one-dimensional insulator with chiral symmetry $ \gamma \tilde{F}(\omega_{\mathrm{ref}},k)\gamma=-\tilde{F}(\omega_{\mathrm{ref}},k)$ with 
$\gamma=\tau_3\otimes \1$, 
the Pauli matrix 
$
\tau_3=
\left(
\begin{array}{cc}
1 & 0 \\
0 & -1 
\end{array}
\right)
$, and the identity matrix $\1$.
Exact zero modes of $\tilde{F}$ are eigenmodes of $F$
\begin{eqnarray}
\label{eq: tildeF tildepsi =0}
&&\tilde{F}(\omega_{\mathrm{ref}},k) |\tilde{\psi}\rangle = 0, \\
\label{eq: tildepsi = psi}
&&
|\tilde{\psi}\rangle
=
\left(
\begin{array}{c}
0  \\
|\psi^{\mathrm{R}}\rangle 
\end{array}
\right)_\tau \ 
\mathrm{or} \ 
\left(
\begin{array}{c}
|\psi^{L}\rangle  \\
0
\end{array}
\right)_\tau,
\end{eqnarray}
with 
$|\psi^{\mathrm{R}}\rangle$ 
($|\psi^{\mathrm{L}}\rangle$)
being right (left) eigenvectors of $F$.

Here, suppose that the winding number $W(\omega_{\mathrm{ref}})$ takes a finite value for a given $\omega_{\mathrm{ref}}\in\mathbb{C}$. Then, the one-dimensional Hermitian insulator hosts edge modes. 
In general, the eigenvalues of the edge modes are small but finite due to the coupling between edges. However, for proper $\omega_{\mathrm{ref}}=\omega_n \in \mathbb{C}$ ($n=1,2,\ldots$), the eigenvalues of $\tilde{F}$ become exactly zero. 
In this case, $\omega_n$ and eigenvectors of $\tilde{F}$ correspond to eigenvalues and eigenmodes of $F$, respectively [see Eqs.~(\ref{eq: tildeF tildepsi =0})~and~(\ref{eq: tildepsi = psi})], which leads to the emergence of skin modes in the presence of nonlinearity.

In the following, we address the characterization of the skin effect for a specific toy model.
We consider a toy model in one dimension which is described by 
\begin{eqnarray}
\label{eq: NL HatanoNelson}
F(\omega,k)&=& 
t_{\mathrm{R}}
\left(
\begin{array}{cc}
0 & e^{\ii k} \\
1 & 0
\end{array}
\right)
+
t_{\mathrm{L}}
\left(
\begin{array}{cc}
0 & 1 \\
e^{-\ii k} & 0
\end{array}
\right)
\nonumber \\
&&
\quad+
\left(
\begin{array}{cc}
\omega^2 &  0\\
0 & \omega^2 +\tanh(\omega)
\end{array}
\right)
\end{eqnarray}
under periodic boundary conditions.
The nonlinearity arises from the last term.

Figure~\ref{fig: NHSE toy}(a) displays the winding number $W(\omega_{\mathrm{ref}})$. 
In the red-colored region, the winding number takes one indicating the emergence of the skin effect.
Indeed, the eigenvalues and eigenmodes exhibit the extreme sensitivity to boundary conditions.
Here, we diagonalize the matrix in the real-space for $L=10$ with $L$ being the number of unit cells.
Figure~\ref{fig: NHSE toy}(b) displays the eigenvalues. 
Under periodic boundary conditions,
eigenvalues form a loop, which is consistent with $W=1$. 
In contrast, under the open boundary conditions, the eigenvalues form lines [Fig.~\ref{fig: NHSE toy}(b)] in the region where $W(\omega_{\mathrm{ref}})$ takes one.
Correspondingly, the eigenmodes also exhibit such sensitivity. 
Under periodic boundary conditions, eigenmodes extend to the bulk. In contrast, under open boundary conditions, they localize around the right edge, which indicates the emergence of skin modes.

The above results demonstrate that the non-Hermitian skin effect survives even under the non-linarity which is predicted by finite values of the winding number $W(\omega_{\mathrm{ref}})$.

\section{
Nonlinearity-induced exceptional points in mechanical systems
}
\label{sec: EPs in MetMat}

 Mechanical systems host nonlinearity-induced symmetry-protected exceptional points and rings which separate stable and unstable modes.   

\subsection{
Spring-mass model of a square lattice}
\label{sec: spring mass}

\begin{figure}[!h]
\begin{minipage}{0.4\hsize}
\begin{center}
\includegraphics[width=1\hsize,clip]{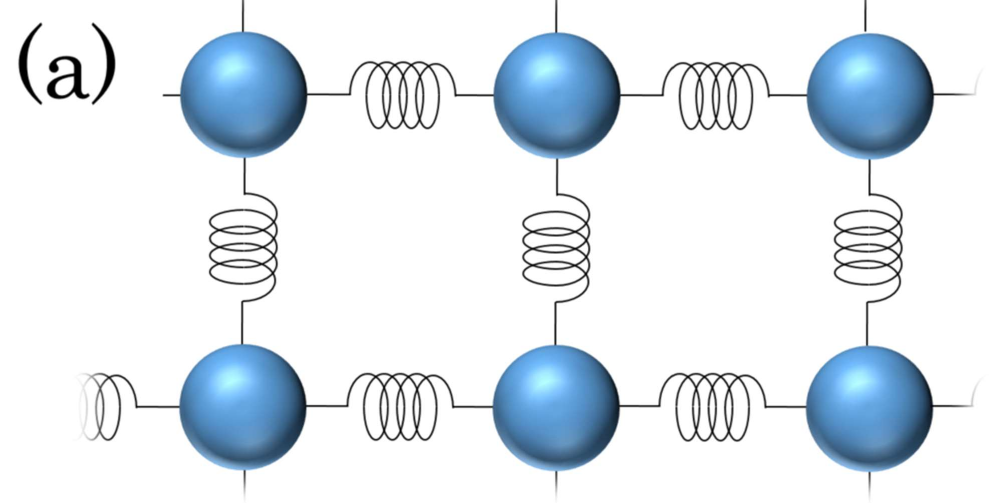}
\end{center}
\end{minipage}
\begin{minipage}{0.48\hsize}
\begin{center}
\includegraphics[width=1\hsize,clip]{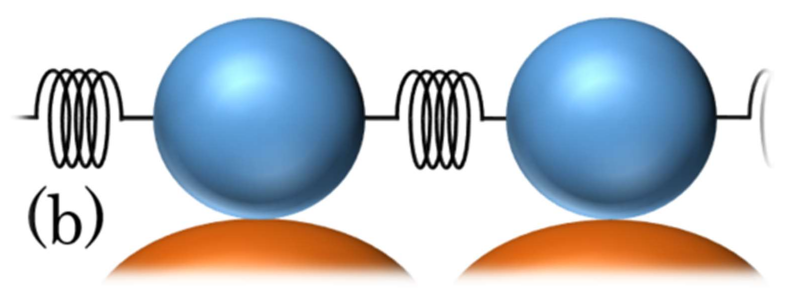}
\end{center}
\end{minipage}
\begin{minipage}{0.48\hsize}
\begin{center}
\includegraphics[width=1\hsize,clip]{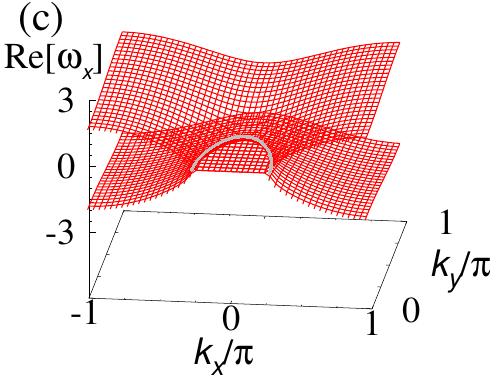}
\end{center}
\end{minipage}
\begin{minipage}{0.48\hsize}
\begin{center}
\includegraphics[width=1\hsize,clip]{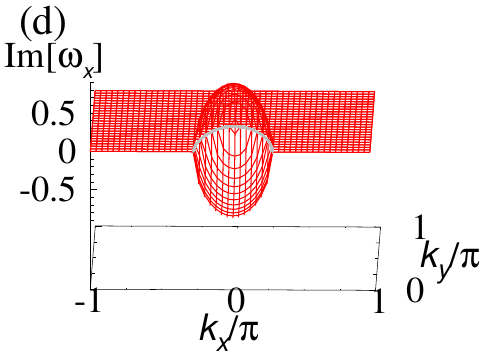}
\end{center}
\end{minipage}
\caption{
(a) and (b): Sketch of the spring-mass model of a square lattice. Neighboring mass points (blue spheres) are connected by springs.
Mass points are on the humps of the floor. (c) [(d)]: The real (imaginary) part of eigenvalues $\omega_{x}$ for $k_y\geq 0$. The data are obtained for $(s_0,v,\eta)=(1,0.7,0.5)$.
The gray lines denote the nonlinearity-induced symmetry-protected exceptional ring.
Eigenvalues $\omega_y$ are obtained by $\pi/2$-rotation in the momentum space.
}
\label{fig: sqlat mech}
\end{figure}

A spring-mass model in two dimensions hosts nonlinearity-induced symmetry-protected exceptional rings. 
Specifically, we consider a spring-mass model of a square lattice.
The equation of motion is written as
\begin{eqnarray}
\label{eq: EOM spring}
-\omega^2 u_\mu (\bm{k})&=& 
-[\xi_\mu(\bm{k})+(1-\eta)\xi_{\bar{\mu}}(\bm{k})-v] u_\mu
\end{eqnarray}
with $\xi_\mu=s_0[2-2\cos(k_\mu)]$ and $\mu=x,y$. The subscript $\bar{\mu}$ takes $y$ ($x$) for $\mu=x$ ($y$).
Here, $u_\mu (\bm{k})$ is Fourier transformed displacement of mass points in the $\mu$ direction. The spring constant is denoted by $s_0\geq 0$. 
The natural length of the spring and the lattice constant are denoted by $l_0$ and $R$, respectively. The ratio between them is denoted by $\eta=l_0/R$.
The potential arising from humps of the floor is described by $v\geq 0$.
The motions  in the $x$ and $y$ direction are decoupled. Thus, oscillation in the $\mu$ direction is described by a nonlinear eigenvalue problem with $1\times 1$ matrix 
\begin{eqnarray}
F_{\mu}(\omega,\bm{k})&=& \omega^2 - [\xi_\mu(\bm{k})+(1-\eta)\xi_{\bar{\mu}}(\bm{k})-v]
\end{eqnarray}
whose eigenvalues are given by
\begin{eqnarray}
\label{eq: omega springmass}
\omega_{\mu}(\bm{k})&=& \pm \sqrt{ \xi_{\mu}(\bm{k})+(1-\eta)\xi_{\bar{\mu}}(\bm{k})-v}.
\end{eqnarray}
This result indicates the emergence of a nonlinearity-induced exceptional ring protected by nonlinear pseudo-Hermiticity [see Eq.~(\ref{eq: UFom*U=Fom})].
Indeed, Figs.~\ref{fig: sqlat mech}(c)~and~\ref{fig: sqlat mech}(d) indicate that both the real and imaginary parts of eigenvalues touch for $\omega_{\mathrm{ref}}=0$.
The zeroth Chern number $N_{0\mathrm{Ch}}$ characterizes the nonlinearity induced exceptional ring; $N_{0\mathrm{Ch}}$ takes $0$ ($1$) inside (outside) of the ring.
The nonlinearity-induced exceptional ring separates stable and unstable modes in the momentum space.
If the eigenvalues $\omega$ have the imaginary part, the corresponding modes are unstable.
For instance, at $\bm{k}=0$, we have $u=
C_- e^{-\omega_0 t} +C_+e^{\omega_0 t} 
$
with $\omega_0=|\omega_\mu(0)|$, $C_\pm = u_0 \pm v_0/\omega_0$, and $u_0, v_0 \in \mathbb{R}$.
Thus, the displacement $u$ diverges, which corresponds to the instability of the translation of all mass points.

\subsection{Kapitza pendulum}
\begin{figure}[!h]
\begin{minipage}{0.4\hsize}
\begin{center}
\includegraphics[width=1\hsize,clip]{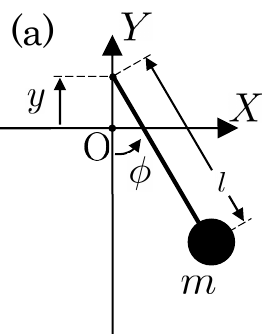}
\end{center}
\end{minipage}
\begin{minipage}{0.58\hsize}
\begin{center}
\includegraphics[width=1\hsize,clip]{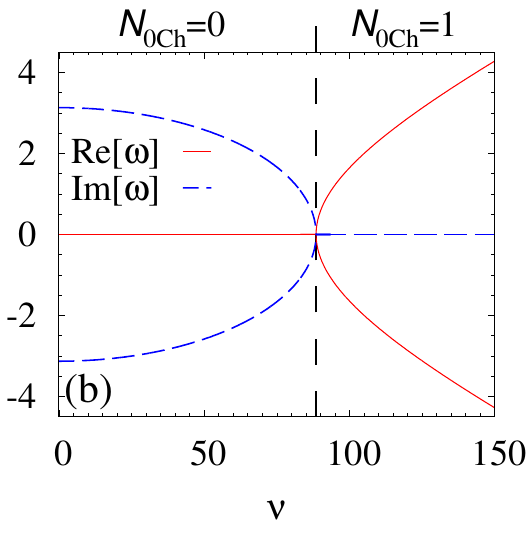}
\end{center}
\end{minipage}
\caption{
(a): Sketch of the Kapitza pendulum. The pivot point oscillates in the $Y$ direction, $y=a\cos (\nu t)$.
(b): Eigenvalues $\omega$ in the one-dimensional parameter space of $\nu \geq 0$. Solid (dashed) lines denote the real (imaginary) part of eigenvalues. The dashed vertical line denotes $\nu_{\mathrm{c}}=\sqrt{2gl}/a$. 
The numerical data are obtained for $(a,g,l)=(0.05,9.8,1)$.
}
\label{fig: Kapitza Pendulum}
\end{figure}

The Kapitza pendulum is another mechanical system hosting nonlinearity-induced exceptional points. 
We consider a pendulum (see Fig.~\ref{fig: Kapitza Pendulum}) where the pivot point oscillates with the angular-frequency $\nu$ and the amplitude $a$. For $\nu>\nu_\mathrm{c}$, the oscillation around $\phi=\pi$ is stable. 
In the one-dimensional parameter space of $\nu$, the nonlinearity-induced symmetry-protected exceptional point specifies the critical value $\nu_\mathrm{c}>0$.

Let $g$, $m$, and $l$ be gravitational acceleration, mass of the mass point, and length of the massless rod.
For $a\ll l$ and $\nu \gg \sqrt{g/l}$, the angle $\phi$ can be decomposed into a slow component $\phi_0$ and a rapid component $\xi$ ($\phi=\phi_0+\xi$).
Here, we focus on the slow component $\phi_0$. 
For $\nu > \nu_{\mathrm{c}}=\sqrt{2gl}/a$, oscillation around $\phi_0 = \pi$ becomes stable.
In this case, the equation of motion of $\theta=\phi_0-\pi$ is written as
\begin{eqnarray}
 \label{eq: EOM pi Kapitza}
 \ddot{\theta} &=& -\frac{g}{l}\left[ \frac{(a\nu)^2}{2gl}-1\right]\theta,
\end{eqnarray}
which indicates that $\theta=\sum_\omega u e^{\ii \omega t}$ satisfies
\begin{eqnarray}
\label{eq: F Kapitza}
F(\omega) u&=& 0,\\
F(\omega) &=& \omega^2 -\frac{g}{l} \left[ \frac{(a\nu)^2}{2gl} -1 \right].
\end{eqnarray}
Equation~(\ref{eq: F Kapitza}) indicates that the oscillation is described by a nonlinear eigenvalue problem.
The eigenvalues $\omega_{\pm}$ are plotted in Fig.~\ref{fig: Kapitza Pendulum}(b) where nonlinearity induces the exceptional point in the one-dimensional parameter space of $\nu$. The symmetry constraint $F^*(\omega)=F(\omega^*)$ [see Eq.~(\ref{eq: UFom*U=Fom})] protects the exceptional point.
On the nonlinearity-induced exceptional point, the zeroth Chern number $N_{0\mathrm{Ch}}$ jumps from $0$ to $1$ with increasing $\nu$ [see Fig.~\ref{fig: Kapitza Pendulum}(b)].
As is the case of the spring-mass system in Sec.~\ref{sec: spring mass}, the modes whose eigenvalues have the imaginary part are unstable.

\section{Summary}
We have elucidated that exceptional points and non-Hermitian skin effects are robust for systems described by the nonlinear eigenvalue problem [Eq.~(\ref{eq: NLEVP})].
The robustness is elucidated by extending the topological invariants to the nonlinear systems.
Notably, even systems without an internal degree of freedom may exhibit exceptional points and their symmetry-protected variants under the nonlinearity, which is in sharp contrast to linear systems.
Such nonlinearity-induced exceptional points are observed in mechanical metamaterials. For instance, the Kapitza pendulum hosts a nonlinearity-induced exceptional point that separates stable and unstable modes in the one-dimensional parameter space.

\section*{
Acknowledgments
}
This work is supported by JSPS KAKENHI Grant Nos.~JP21K13850, 23K25788 and JP23KK0247, JSPS Bilateral Program No.~JPJSBP120249925.
T.Y is grateful for the support from the ETH Pauli Center for Theoretical Studies and the Grant from Yamada Science Foundation.


\begin{thebibliography}{107}
\bibitem [{gap()}]{gaplinear_ftnt}%
  \BibitemOpen
  \href@noop {} {}\bibinfo {note} {{ For linear systems, it is defined that the
  point-gap opens for a given $\omega_{\mathrm{ref}}\in\mathbb{C}$ when
  $\mathrm{det}[H(\bm{k})-\omega_{\mathrm{ref}}\1 ]\neq 0 $ is satisfied
  }}\BibitemShut {NoStop}%
\bibitem [{EPL()}]{EPLinear_ftnt}%
  \BibitemOpen
  \href@noop {} {}\bibinfo {note} {Recalling the relation $b(\bm{k})=k_x+\ii
  k_y$, we can see that Eq.~(\ref{eq:w+, w- gen}) is reminiscent of the
  dispersion relation of an exceptional point described by a linear eigenvalue
  problem of $2\times 2$ non-Hermitian matrix. For instance, consider the
  non-Hermitian matrix $ H(\bm{k}) = \left( \begin{array}{cc} 0 & k_x+\ii k_y
  \\ -1 & 0 \end{array} \right). $ Diagonalizing this matrix, we obtain
  $\omega_\pm(\bm{k})=\pm\sqrt{k_x+\ii k_y}$ \hspace{-2mm}}\BibitemShut
  {NoStop}%
\end{thebibliography}
%


\appendix

\section{Derivation of Eq.~(\ref{eq: EOM spring})}
\label{sec: EOM spring app}
For the system illustrated in Fig.~\ref{fig: sqlat mech}(a), the potential arising from springs is written as
\begin{eqnarray}
U_{\mathrm{s}}&=& \frac{s_0}{2} \sum_{\langle ij\rangle}
\left[
2(1-\eta) \delta\bm{R}_{ij}\cdot \delta \bm{u}_{ij}  +(1-\eta) \delta \bm{u}_{ij}\cdot \delta \bm{u}_{ij} \right. \nonumber \\
&&
\left.
\quad\quad +\eta R^{-2} (\delta\bm{R}_{ij}\cdot \delta \bm{u}_{ij})(\delta\bm{R}_{ij}\cdot \delta \bm{u}_{ij})
\right]
\end{eqnarray}
with $\eta=l_0/R$, $\delta \bm{u}_{ij}=\bm{u}_{i}-\bm{u}_{j}$, and $\delta \bm{R}_{ij}=\bm{R}_{i}-\bm{R}_{j}$.
Vector $\bm{u}_{i}(t)$ describes the displacement of mass point at time $t$ and site $j$ whose $\mu$-component $u_{i\mu}(t)$ denotes the displacement in the $\mu$ direction ($\mu=x,y$).
Vector $\bm{R}_{i}$ specifies the position of site $i$.
The natural length of the spring and the lattice constant are denoted by $l_0$ and $R$, respectively.
The summation is taken over all of the nearest neighboring sites.
Taking into account the potential arising from humps of the floor $U_{\mathrm{f}}=\frac{v}{2}\sum_{i} \bm{u}^2_i$ ($v\geq 0$), we obtain the equation of motion
\begin{eqnarray}
\label{eq: EOM spmass app}
-\omega^2 
\left(
\begin{array}{c}
u_{x}(\omega,\bm{k})  \\
u_{y}(\omega,\bm{k})
\end{array}
\right)
&=&
-[M_{\mathrm{s}}(\bm{k})-M_{\mathrm{f}}]
\left(
\begin{array}{c}
u_{x}(\omega,\bm{k})  \\
u_{y}(\omega,\bm{k})
\end{array}
\right)
\nonumber\\
\end{eqnarray}
with
\begin{eqnarray}
M_{\mathrm{s}}(\bm{k}),
&=&
\left(
\begin{array}{cc}
\xi_x(\bm{k})+(1-\eta)\xi_y(\bm{k}) &  0 \\
0 & (1-\eta)\xi_x(\bm{k})+\xi_y(\bm{k})
\end{array}
\right), \nonumber \\ 
M_{\mathrm{f}}
&=&
v
\left(
\begin{array}{cc}
1 &  0 \\
0 & 1
\end{array}
\right), \nonumber \\
\xi_{\mu}(\bm{k}) &=& 2-2\cos k_\mu. \nonumber 
\end{eqnarray}
Here, we have applied the Fourier transformation
\begin{eqnarray}
u_{j\mu}(t) &=& \frac{1}{\sqrt{N_{\mathrm{site}}}}\sum_{\bm{k}} \int \! \frac{d\omega}{\sqrt{2\pi}}  e^{i(\omega t+ \bm{k}\cdot\bm{R}_j)} u_{\mu}(\omega,\bm{k}). \nonumber \\
\end{eqnarray}
Equation~(\ref{eq: EOM spmass app}) is nothing but Eq.~(\ref{eq: EOM spring}) because matrices $M_\mathrm{s}$ and $M_{\mathrm{f}}$ are diagonal.

\section{Derivation of Eq.~(\ref{eq: EOM pi Kapitza})}
The Lagrangian of the pendulum [see Fig.~\ref{fig: Kapitza Pendulum}] is given by 
\begin{eqnarray}
\mathcal{L}&=& \frac{m}{2}(l^2\dot{\phi}^2 +2l\dot{\phi}\dot{y}\sin \phi+\dot{y}^2)+mg(l\cos \phi -y) \nonumber \\
\end{eqnarray}
with $y=-a\cos(\nu t)$ and $\dot{\phi}=\frac{d}{dt}\phi$.
The Lagrangian $\mathcal{L}$ is rewritten as
\begin{eqnarray}
\mathcal{L} &=& \mathcal{L}' +\frac{d}{dt}(\dot{y}\cos \phi), \\
\mathcal{L}' &=& \frac{m}{2}(l^2\dot{\phi}^2 +2l \ddot{y}\cos \phi ) +mgl\cos \phi,
\end{eqnarray}
with $\ddot{y}=\frac{d^2}{dt^2}y$ and $\frac{d}{dt}$ denoting derivative with respect to time.
Thus, the equation of motion 
$
\frac{d}{dt} \frac{\partial \mathcal{L}' }{\partial \dot{\phi} } = \frac{\partial \mathcal{L}' }{ \partial \phi }
$
is written as
\begin{eqnarray}
\label{eq: EOM of phi app}
\ddot{\phi} &=& -\frac{1}{l}(\ddot{y}+g)\sin \phi.
\end{eqnarray}

Suppose that the amplitude and frequency of the pivot point satisfy $ a \ll l$ and $\nu \gg \sqrt{g/l}$.
In this case, $\phi$ can be decomposed into a slow component $\phi_0$ and a rapid component $\xi$ ($\phi=\phi_0+\xi$).
The rapid component $\xi$ is due to oscillation of the pivot point and thus is written as
\begin{eqnarray}
\label{eq: xi = -y/l sin phi0 app}
\xi &=& -\frac{y}{l}\sin \phi_0.
\end{eqnarray}
Substituting Eq.~(\ref{eq: xi = -y/l sin phi0 app}) into Eq.~(\ref{eq: EOM of phi app}), we obtain
\begin{eqnarray}
\ddot{\phi}_0 &=& -\frac{1}{l}(g+\ddot{y}) \left( \sin \phi_0 -\frac{y}{l}\sin \phi_0 \cos \phi_0 \right) -\ddot{\xi}.
\end{eqnarray}
Here, we have used $\sin \xi \sim \xi$ and $\cos \xi \sim 1$, supposing that $a\ll l$ holds.

Taking the average of the rapid component, we have the equation of motion of $\phi_0$
\begin{eqnarray}
\ddot{\phi}_0 &=& -\frac{\partial V}{\partial \phi_0}, \\
V &=& -\frac{g}{l} \left[\cos \phi_0 -\frac{(a\nu)^2}{4gl} \sin^2\phi_0 \right].
\end{eqnarray}
The effective potential $V$ has two minima for $\nu > \sqrt{2gl}/a$; $\phi_0=0$ and $\pi$.

Thus, for $\theta=\phi_0-\pi \ll 1$, we have
\begin{eqnarray}
\ddot{\theta} &=& -\frac{g}{l} \left[ \frac{(a\nu)^2}{2gl}-1 \right] \theta,
\end{eqnarray}
which is nothing but Eq.~(\ref{eq: EOM pi Kapitza}).

\end{document}